# JAMMING AND THE ONSET OF GRANULATION IN A MODEL PARTICLE SYSTEM

Daniel J. M. Hodgson, Michiel Hermes & Wilson C. K. Poon

School of Physics and Astronomy, James Clerk Maxwell Building, The University of Edinburgh, Peter Guthrie Tait Road, Edinburgh, EH9 3FD, UK
d.j.m.hodgson@sms.ed.ac.uk

## ABSTRACT

Granulation is a ubiquitous process crucial for many products ranging from food and care products to pharmaceuticals. Granulation is the process in which a powder is mixed with a small amount of liquid (binder) to form solid agglomerates surrounded by air. By contrast, at low solid volume fractions $\phi$, the mixing of solid and liquid produces suspensions. At intermediate $\phi$, either granules or dense suspensions are produced, depending on the applied stress. We address the question of how and when high shear mixing can lead to the formation of jammed, non-flowing granules as $\phi$ is varied. In particular, we measure the shear rheology of a model system - a suspension of glass beads with an average diameter of $\approx 7$ µm - at solid volume fractions $\phi \gtrsim 0.40$. We show that recent insights into the role of inter-particle friction in suspension rheology allow us to use flow data to predict some of the boundaries between different types of granulation as $\phi$ increases from $\approx 0.4$ towards and beyond the maximum packing point of random close packing.

## 1. INTRODUCTION

The process of granulation involves applying high-shear to high-solid-content particle-liquid mixtures. The study of the controlled deformation of suspensions and other materials constitutes the science of rheology. A priori, therefore, one might expect that suspension rheology should contribute significantly to the understanding of the granulation process. In practice, however, there is surprisingly little literature making this link. An important reason for this may be that the rheology of suspensions with particles in the relevant size range, i.e. between the colloidal and granular regimes ($\gtrsim 1$ µm and $\lesssim 20$ µm), is not at all well understood. In particular, shear thickening is ubiquitous for such suspensions. A number of recent advances [1–3] have led to a new theoretical framework for understanding shear thickening at a particle level, and the framework has been applied to interpret the rheology of a model suspension with particles in the intermediate size regime [4]. In this paper, we present an experimental rheological study of a model system of glass beads in the same size regime that shows clear shear-thickening. Crucially, our model system also readily granulates. This enables us to explore how suspension rheology controls granulation.

Our work is motivated by experimental observations from both granulation and suspension rheology. Previous work in granulation investigating the role of solid volume fraction, $\phi$, reported that beyond a certain liquid content (i.e. as $\phi$ is lowered) for a given system, granulation ceases and the system becomes wet, tending towards a suspension with a high yield stress (a paste or slurry) [5,6]. In other words, granulation ceases as $\phi$ is lowered to the point when suspension rheology begins. Conversely, suspension rheology experiments at high volume fractions often stop at a certain volume fraction, beyond which sample preparation becomes impossible as the sample breaks up and fractures under normal handling stresses. In other words, granulation starts as $\phi$ is increased at the point where the viscosity $\eta$, diverges and suspension rheology becomes impossible. The theoretical link between these two areas of phenomenology has been pointed out before [7,8], but there has so far been little experimental follow up.

Within the granulation literature, it is known that three distinct regimes are found as the solid volume fraction $\phi$ is increased from $\approx 0.4$ [9]. From rheological measurements of our model system, we find that the transitions from one mixing regime to another coincide with two successive stress-dependent divergences in the suspension viscosity. The first divergence is associated with flow dominated by inter-particle frictional contacts, while the second, higher-$\phi$, divergence is associated with flow dominated by inter-particle (hydrodynamically) lubricated contacts. Conceptually, this observation provides us with new insights into the mechanism that drives the formation of granules; practically, the same observation leads to protocols for predicting granulation regimes from rheological measurements alone. Traditionally, demarcating the different granulation regimes is typically done by trial and error, with associated time and material costs. In comparison, our rheological protocol is not only more efficient in effort and samples, but also underpinned by theoretical understanding, so that the output of the protocol already gives significant clues to how the system may be 'tuned'. Since $\phi$ constitutes one of the major axes in granulation parameter space, our results therefore constitute an important advance in predictive granulation.

### 1.1 Shear Thickening

To provide a context and the necessary background for our work, we first review recent insights into the role of friction in suspension flow; in particular, how friction controls shear thickening. Recent experiments [4] have found that in a system of sterically-stabilised spherical particles with diameters in the range 1 μm $\lesssim D \lesssim$ 10 μm, the viscosity *vs* volume fraction plot, $\eta(\phi)$, exhibits two branches (cf. Figure 4). At low enough applied stresses, there is a lower branch, $\eta_1(\phi)$, which can be fitted to a Krieger-Dougherty form:

$$\eta = \eta_0 \left(1 - \frac{\phi}{\hat{\phi}}\right)^{-\lambda}, \hspace{4cm} Eq\ (1)$$

that diverges at $\hat{\phi} \approx 0.64$, which is the point of random close packing (RCP) for monodisperse spheres. However, beyond a particle-size-dependent onset stress $\sigma^*$, the viscosity rises continuously to an upper newtonian plateau. The plateau value, $\eta_2(\phi)$, can also be fitted to Equation 1, but now diverging at a lower volume fraction: for monodisperse spheres $\hat{\phi} \approx 0.56$.

Recent theory [1] and simulations [2,3] predict precisely such a two-branch structure. Theory suggests and simulations show that the divergence of $\eta_2(\phi)$ is controlled by inter-particle friction, $\mu$. If the inter-particle friction coefficient of monodisperse hard spheres is set to zero, $\mu = 0$, then $\eta_2(\phi)$ also diverges at $\phi_{\text{RCP}} \approx 0.64$. As $\mu$ increases from zero, the divergence of $\eta_2(\phi)$, which from now on we will label as $\phi_m$, decreases from 0.64, reaching $\phi_m = 0.56$, as $\mu \to \infty$. Shear thickening occurs at an onset stress, $\sigma^*$ which is the stress needed to overcome inter-particle repulsion and push them into frictional contact. Shear thickening is associated with the suspension making a transition from the lower viscosity branch to the upper viscosity branch as an increasing proportion of contacts become frictional.

Within this framework, it is clear that some form of flow catastrophe occurs beyond $\phi_m$, the point at which the viscosity of particles flowing with frictional contacts diverges. Beyond $\phi_m$, there is no 'upper branch' to which the suspension in the 'lower branch' could make a transition at $\sigma > \sigma^*$. Experimentally, it is observed that samples in this regime undergo discontinuous shear thickening, and shows edge instabilities and fracture in the rheometer geometry [4,7]. This is probably associated with the capillary pressure at the sample-air interface no longer being able to confine the particles against the pressure to dilate [7,10]. It has been suggested before that such rheological phenomena were related to granulation [7]. We therefore proceed to test this suggestion experimentally in the light of recent advances in understanding shear thickening.

## 2. MATERIALS AND METHODS

Industrial formulations invariably consist of multiple components, for example active drug and excipients in a pharmaceutical product [11]. However these components have diverse physical properties such as size distribution, shape or degree of wetting in the liquid phase [12]. Such complexities can mask the underlying physics. We therefore use the simplest experimental system that still exhibits the physical behaviour of interest, namely shear thickening and granulation: a suspension of a single type of solid particles in a simple liquid binder.

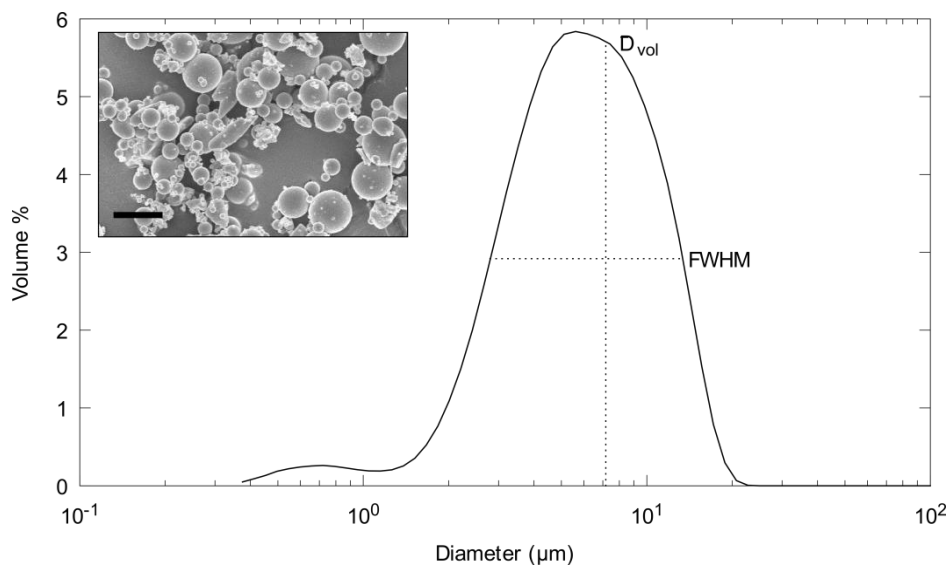

Figure 1. Volume-weighted size distribution of the Spheriglass 5000 soda-lime glass spheres used in this work. Size was measured using static light scattering. Volume weighted average diameter ($\overline{D}_{vol}$) is shown (vertical dashed line). The full-width half maximum (horizontal dashed line) is used to provide an estimate of the volume weighted polydispersity, though this neglects the large number of small particles in the secondary peak at ≈ 0.7 μm. (Inset) Scanning electron micrograph of Spheriglass 5000. Particles are mainly polydisperse spheres with some shards. Scale bar is 10 μm.

### 2.1 Materials

Soda-lime glass microspheres (Potters Spheriglass A-Type 5000) were used for the solid phase throughout this work. The powder was dried at 120°C for at least 3 hours prior to either granulation or preparation for rheology in order to remove any moisture. Static light scattering (Beckman Coulter LS 13 320) was used to determine the particle size distribution. The volume weighted size distribution is shown in Figure 1, with the mean diameter, $\overline{D}_{vol}$ = 7.16 μm. The distribution shows a high level of polydispersity (p.d. of main peak ≈ 147%) calculated from the full-width half-maximum (FWHM) divided by $\overline{D}_{vol}$. The wide size distribution is typical in industrial formulations (for example [13,14]).

A mixture of glycerol (Fisher Chemical, 90% by volume) and distilled water (10% by volume) was used as the liquid phase. This combination was chosen for a number of reasons: a high enough viscosity was achieved to suppress sedimentation during rheology experiments, but still with a high enough water content to minimise water absorption during the time scale of our experiments. The glycerol-water mixture is also polar enough that our particles are charged and repulsive.

In this work $\phi$ is calculated by measuring the mass of each component, then calculating its volume from the measured material density ($\rho_{glass} = 2.46$ g cm$^{-3}$, $\rho_{binder} = 1.24$ g cm$^{-3}$). Absolute volume fractions are expected to be accurate to within a few percent of the true value, with relative volume fractions achieving much better accuracy.

### 2.2 Granulation

Granulation was carried out in a custom-built granulator consisting of a cylindrical glass dish (diameter 110 mm) with an overhead uniaxial impeller (see Figure 2). The impeller consists of three identical blades with 120° angular separation and a front edge height of 17.9 mm at a rake of 45°. Liquid binder is added from above the powder bed at a single point located at a radial distance of 32 mm from the central axis through tubing with an inside diameter of 1.5 mm.

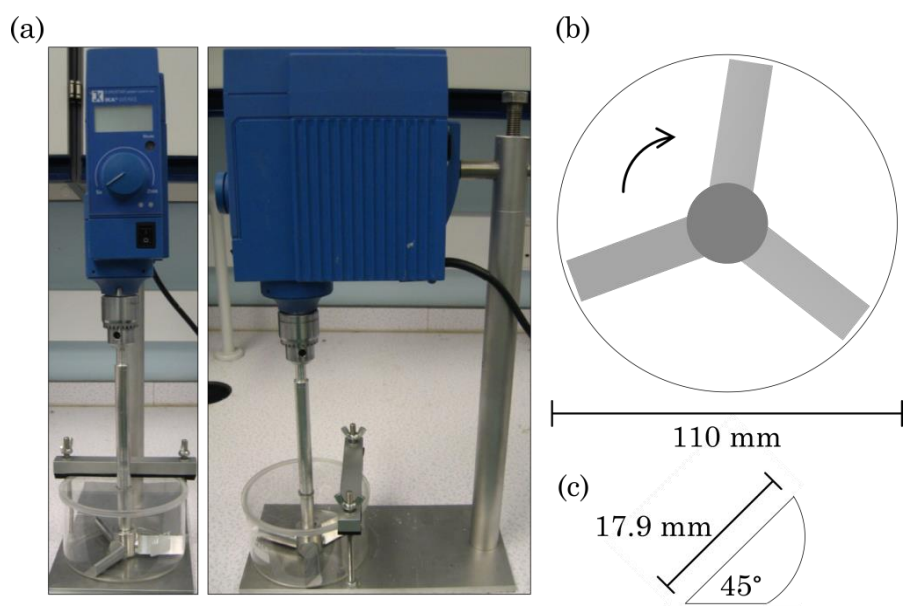

Figure 2. (a) The granulator comprises a rate-controlled over-head mixer that drives a uniaxial impeller with 3-blades at speeds between 50-2000 rpm. (b) Granulation takes place in a cylindrical glass dish. The wall clearance between the impeller blades and dish is < 1 mm. Liquid binder is added at a single point to the top of this container by way of syringe pump. (c) A cross-section of the impeller blades. The bottom of the blades sit < 1 mm above the bottom of the container and have a rake of 45°.

The solid phase was added first and pre-mixed at ≈ 500 rpm for 30 seconds. A fixed volume of the liquid phase was then added at a rate of 15 mL min$^{-1}$ using a syringe pump. Samples were mixed with an impeller speed of ≈ 500 rpm for between 3 and 6 minutes depending on the state of the system following liquid addition.

Samples were then removed from the granulator and sealed in vials. The initial state following mixing was recorded (see Figure 5). A sub-sample was then separated before applying low shear on a vortex mixer for 90 seconds in order to observe whether a low stress flowing state could be achieved. Samples that did not flow after application of low shear on the vortex mixer were then slowly rolled for longer periods of time (> 72 hours) to observe further structural evolution over longer periods of time. Samples of the same volume fraction that flowed on the vortex mixer were also rolled for the same period for comparison.

## 2.3 Rheology

Samples for rheology were prepared by taking a sample at $\phi = 0.60$ prepared using the method described in Section 2.2 and diluting with a glycerol-water mixture to the target $\phi$ and left on a roller mixer for at least 72 hours. Samples were carefully sealed to minimise moisture ingress.

An Anton-Paar stress-controlled rheometer (MCR-301) was used with sand-blasted 50 mm parallel plates in order to minimise slip. We worked at a gap height of 1 mm, and we trimmed samples at a height of 1.025 mm in order to create a reproducible edge meniscus. Stress sweeps were performed. Samples were equilibrated at each stress to give the steady state viscosity, which was measured for a downwards and upwards stress sweeps and then averaged. Experimental time was kept to a minimum in order to reduce the effects of sedimentation and hydration of the binder through atmospheric moisture absorption, which both lead to a decrease in the viscosity.

## 3. RESULTS AND DISCUSSION

### 3.1 Rheology

The average steady state viscosity as a function of applied stress is shown in Figure 3. These samples in the range $0.40 \leq \phi \leq 0.55$ exhibit clear continuous shear thickening behaviour. The viscosity at each $\phi$ increases continuously from a low-shear plateau to a high-shear plateau starting at a critical onset stress ($\sigma^* \approx 0.8$ Pa). At the highest stresses reached (data not shown), the measured viscosity decreases; this can be due to several factors, for example slip, particle migration and loss of material from edge fracture [15,16].

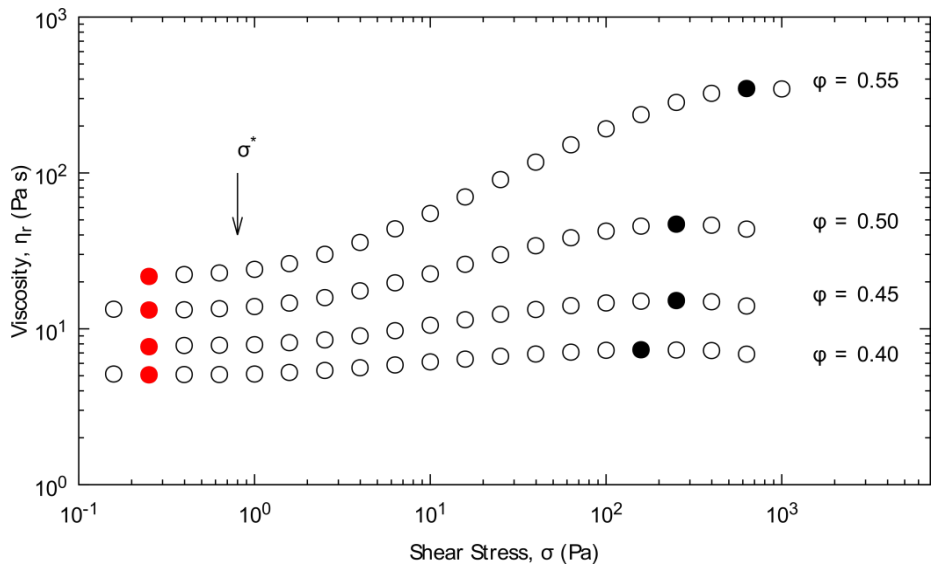

Figure 3. Steady state viscosity as a function of applied stress. Clear shear thickening behaviour is displayed, with the viscosity increasing from a minimum below the onset stress, $\sigma^*$ (values given by ● are plotted as $\eta_1(\phi)$ in Figure 4), up to a maximum value (values given by ● are plotted as $\eta_2(\phi)$ in Figure 4). Beyond the highest-stress data point in each case, we observe instabilities and fracture.

The viscosities at the two Newtonian plateaus are plotted as a function of volume fraction in Figure 4, which clearly fall onto two branches with separate points of incipient divergence. Using a least-squares fit of the Kriger-Dougherty model, Equation 1, we find that the lower branch, $\eta_1(\phi)$, diverges at $\phi_m = 0.572$ (with $\lambda = 1.805$), and the upper branch, $\eta_2(\phi)$, diverges at a higher volume fraction that we identify as random close packing for our system: $\phi_{RCP} = 0.711$ (with $\lambda = 2.078$). These observations correspond closely to the picture predicted by theory [1,17] and simulations

[2,3], and observed in experiments using sterically-stabilised polymer spheres of a similar size [4]. The lower branch corresponds to flow of our charged-stabilised glass spheres with lubricated inter-particle contacts, while the upper branch corresponds to shear-thickened flow with frictional inter-particle contacts. Below the onset stress $\sigma^*$, particles are separated by a repulsive barrier, which in our system is due to surface charge. As the applied stress increases above $\sigma^*$, an increasing proportion of particles are pushed into frictional contact.

At $\phi < \phi_m$ the system can transition smoothly between the lower viscosity branch to the shear-thickened, higher viscosity branch. However, beyond $\phi_m$ there is no stable shear thickened state to which the system can transition. Once the particles are driven into frictional contact when the stress reaches $\sigma^*$, the system jams. In order to resume flow, the system must dilate [7].

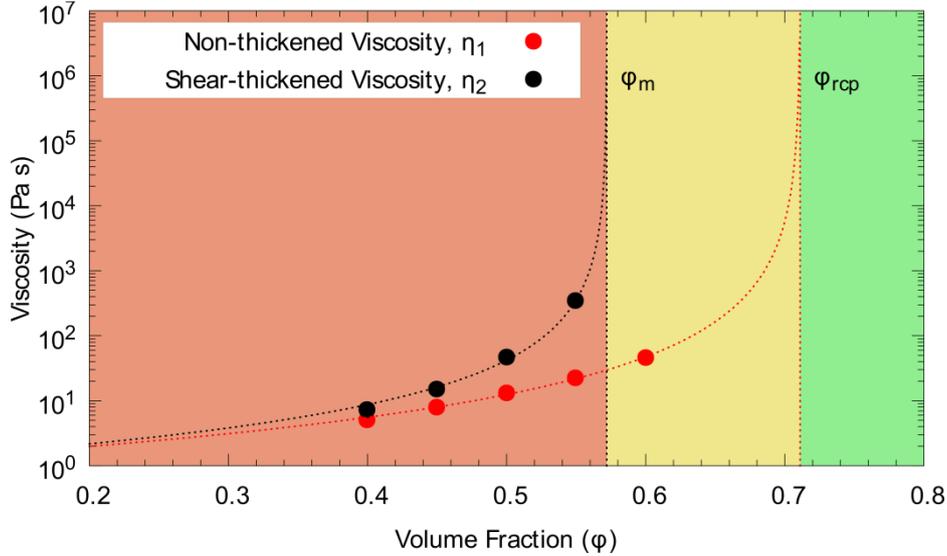

Figure 4. A plot of the shear-thickened viscosity (●) and high shear viscosity (●) as a function of volume fraction with their fits to the Krieger-Dougherty relation diverging at $\phi_m$ and $\phi_{RCP}$ respectively. At low volume fractions $\phi < \phi_m$) the system can flow at all applied stresses. At intermediate volume fractions ($\phi_m < \phi < \phi_{RCP}$) the system can only flow for applied stresses below $\sigma^*$. At higher stresses, the system jams. Samples prepared with high volume fractions ($\phi > \phi_{RCP}$) are unable to flow at any applied stress and the system is jammed at all applied stresses. In order to flow the system must dilate.

For a suspension, such dilation means deforming the containing surface, so that particles can be partially pushed out. The stress scale, $\sigma_b$, required for this is to happen must scale as the ratio of the surface tension, $\Sigma$, and the particle diameter, $D$, and has been experimentally determined to follow $\sigma_b \approx 0.1 \Sigma/D$ [10], although the exact value depends on several factors such as contact angle, geometric factors and particle surface roughness. For our system, $\Sigma \approx 64$ mN m$^{-1}$ [18] and using the mean diameter, $\overline{D}_{vol} = 7.16$ μm, gives a stress scale of $\sigma_b \approx 900$ Pa, which does correspond to the stress beyond which the $\phi = 0.55$ sample starts to fracture, with lower-$\phi$ samples exhibiting a maximum in viscosity at lower stresses. This could be attributed to the effects of slip, which should decrease with increasing $\phi$, since the particle pressure has been found to increase as $\phi \to \phi_m$ [19], which could lead to better purchase on the rheometer geometry and consequently less slip.

### 3.2 Granulation

Granulation experiments reveal a range of behaviour depending on the volume fraction (see Figure 5). Below $\phi = 0.60$ samples rapidly form a suspension and remain in this state. For intermediate volume fractions ($0.60 \lesssim \phi \lesssim 0.70$), the system is unable to flow at high stresses and can form granules. However, these granules are metastable and can flow under low shear. At volume fractions above $\phi \gtrsim 0.70$ the system granulates and can no longer form a flowing suspension at any applied stress.

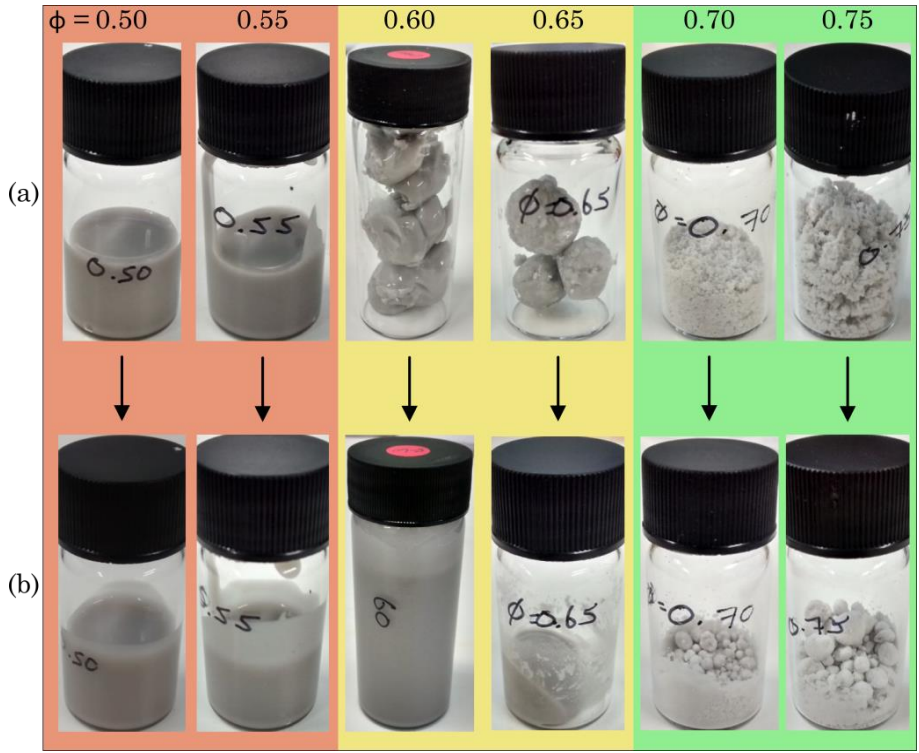

Figure 5. The final states of high-shear mixing as a function of volume fraction. (a) The state of the sample immediately after high shear mixing. (b) The state of samples after gentle agitation (low stress). ■ Volume fractions below ≈ 0.60 exit the granulator as a flowing suspension and remain in that state. ■ Samples prepared in the range $0.60 \lesssim \phi \lesssim 0.70$ are solid upon exiting the granulator, but are able to flow following application of low stresses. ■ Samples prepared with $\phi \gtrsim 0.70$ exit the high shear mixer as dry granules and remain in that state indefinitely.

Our findings suggest that the transition from suspension formation to stress-dependent granulation to permanent granulation, Figure 5, is related to the two viscosity branches shown in Figure 4. Below $\phi_m$, high-shear agitation lead to the formation of a flowing suspension. In the range $\phi_m < \phi < \phi_{RCP}$, the system forms granules that, however, liquify upon the application of low stresses. These granules are kept in frictional contact by the surface tension, once these frictional contacts are broken by small oscillatory forces these granules transition back to the frictionless branch. Above $\phi_{RCP}$, permanent, dry granules are formed.

Our findings may be compared with previous studies. Limited work [5,6] using a mixer torque rheometer into varying liquid content have found a change in granulation behaviour at a viscosity maximum, but the link with particle interactions and rheology was not made. Recall that we also find such maxima: the curves start to decrease at high stresses beyond the last data points shown in Figure 3; but we do not show such data because they are due to slip and other specific experimental effects. There are also similarities between our findings and interpretation with a previous study [7].

## 4. PRACTICAL IMPLICATIONS FOR GRANULATION

We now summarise the implications of our findings for the practice of granulation. First, our comparative study of rheology and granulation suggests that the onset of granulation at increasing $\phi$ can be predicted from a limited set of rheological measurements to give $\phi_m$ and $\phi_{RCP}$. As a bonus, such measurements only require small amounts ($\lesssim 2$ mL) of material per sample.

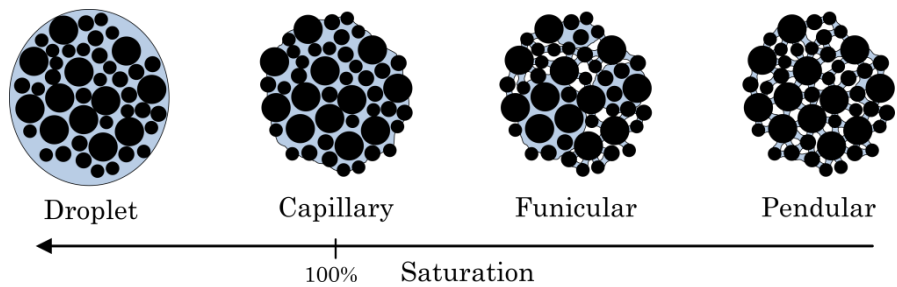

Figure 6. The different types of granule as a function of saturation (after [5,9]). Volume fraction increases from left to right (decreasing saturation). The droplet state is essentially a suspension state that can flow and describes the state of a system with $\phi < \phi_m$ and the low stress state of systems with $\phi_m \lesssim \phi < \phi_{RCP}$. The capillary state represents the high stress state for systems with $\phi_m \lesssim \phi < \phi_{RCP}$ and corresponds to 100% saturation. The final two states, funicular and pendular, represent the state of the system with $\phi > \phi_{RCP}$. In these states there is no longer enough liquid to fully saturate the granule and air is included in the structure. These type of granules are permanent and cannot achieve a flowing suspension state.

Once determined, the locations of $\phi_m$ and $\phi_{RCP}$ provide information about the type of granules that will be produced at a given $\phi$. Below $\phi_m$, granulation is not possible. A system prepared with $\phi_m \lesssim \phi < \phi_{RCP}$ will produce transient granules upon being sheared at high enough stresses. These granules will be fully saturated and, provided the particle interaction is repulsive, they are able to melt and flow on the application of low stresses. $\phi_{RCP}$ is the most efficient possible amorphous packing, and adding even less liquid will result in less than 100% pore saturation and the inclusion of air. Thus, moving beyond $\phi_{RCP}$ will move granules from the capillary state to the funicular and pendular states, Figure 6.

Within this framework, influencing the way granulation depends on $\phi$ becomes a matter of 'tuning' $\phi_m$ and $\phi_{RCP}$. The two divergences are determined by the packing efficiency of a given system. By tuning the packing efficiency through particle shape and size distribution [20,21] it should be possible to change $\phi_m$ and $\phi_{RCP}$ to attain a range of desired onsets of different types of granulation. Thus, for example, it is known that using a binary mixture increases $\phi_{RCP}$ [20]. $\phi_m$ should also increase with $\phi_{RCP}$ though this effect has not been specifically studied. Recent simulations show that $\phi_m$ is very sensitive to the inter-particle friction coefficient. There has been a long history of 'tuning' this parameter using a range of additives [22], opening up the possibility for their use in controlling granulation.

For new formulations, simply measuring mass fractions of solid and liquid components is not enough to predict where a system will begin to granulate, and makes comparison to other systems with different size distributions and surface properties difficult. Instead, measuring $\phi_m$ and $\phi_{RCP}$ allows different systems to be easily compared and predict from these two values how a new formulation will behave before the first trial granulation experiment.

## 5. CONCLUSIONS

We have correlated the rheology of a suspension of $\approx 7$ µm charge-stabilised glass beads to their granulation behaviour. The rheology confirms recent models of shear thickening in such suspensions [1–3,17], and corresponds closely to observations made in similar-sized sterically-stabilised polymer colloids [4]. The viscosity as a function of solid volume fraction, $\eta(\phi)$, displays two branches. A lower, lubricated, branch diverges at random close packing, $\phi_{RCP}$, while a higher, frictional, branch diverges at the system's 'jamming point', $\phi_m < \phi_{RCP}$. The two points correspond to previously-known transitions in granulation behaviour: from non-granulating to low-yield-stress granules at $\phi_m$, and from the latter to under-saturated, permanent granules at $\phi_{RCP}$. Previously, these transitions have to be identified using trial and error granulation experiments; our results show how they can be identified by rheology using small amounts of samples. Relating these transitions to the physics of particle packing and inter-particle frictional interaction also suggests how to 'tune' them. Our findings show that suspension rheology and granulation are closely related, and further work along this direction is clearly warranted.

**LIST OF SYMBOLS**

| Symbol | Description | Units |
|---|---|---|
| $\overline{D}_{vol}$ | Average volume weighted diameter | [m] |
| $\eta$ | Viscosity | [Pa s] |
| $\eta_0$ | Viscosity of suspending liquid | [Pa s] |
| $\eta_r$ | Relative viscosity; $\eta_r = \frac{\eta}{\eta_0}$ | [-] |
| $\mu$ | Coefficient of inter-particle static friction | [-] |
| $\sigma$ | Stress | [Pa] |
| $\sigma^*$ | Onset stress for shear thickening | [Pa] |
| $\Sigma$ | Surface tension | [mN m$^{-1}$] |
| $\phi$ | Volume fraction; $\phi = \frac{V_{solid}}{V_{solid}+V_{liquid}}$ | [-] |
| $\phi_m$ | Volume fraction at which 'upper branch' diverges | [-] |
| $\phi_{RCP}$ | Volume fraction of maximum amorphous packing density | [-] |

**ACKNOWLEDGEMENTS**

DJMH held an EPSRC studentship. MH and WCKP were funded by EPSRC grant EP/J007404/1. We thank Ben Guy and Mike Cates for illuminating discussions.**REFERENCES**